\documentclass[useAMS,usenatbib,usegraphicx]{mn2e}

\usepackage{journals}
\usepackage{multirow}
\usepackage{url}
\usepackage{threeparttable}
\usepackage{graphicx}
\usepackage{longtable}
\usepackage[varg]{txfonts}
\usepackage{bm}
\usepackage{color}
\usepackage{ulem}

\def\ergs{\mbox{erg~s$^{-1}$}}

\def\apj{ApJ}
\def\apjl{ApJ}
\def\apjs{ApJS}
\def\aa{A\&A}
\def\aap{A\&A}

\def\sa{SvA}

\def\mnras{MNRAS}

\def\pasj{Publ. Astron. Soc. Japan}
\def\memsai{Mem. Soc. Astron. It.}

\def\nat{Nature}

\def\physrep{Phys.Rep.}

\def\beq#1{\begin{equation}\label{#1}}
\def\eeq{\end{equation}}
\def\beqa#1{\begin{eqnarray}\label{#1}}
\def\eeqa{\end{eqnarray}}

\def\Eq#1{Eq.~(\ref{#1})}

\def\myfrac#1#2{\left(\frac{#1}{#2}\right)}

\def\comment#1{\relax}

\title[Spin-up/spin-down in GX\,304-1]{Spin-up/spin-down of neutron star in Be-X-ray binary system GX\,304-1}
\author[Postnov et al.]{Postnov K.A.$^{1}$\thanks{E-mail:pk@sai.msu.ru}, Mironov A.I.$^{2,3}$,
Lutovinov A.A.$^{2}$, Shakura N.I.$^{1}$, Kochetkova A.Yu$^{1}$,
\newauthor Tsygankov S.S.$^{4}$\\
$^{1}$ Moscow M.V. Lomonosov State University, Sternberg Astronomical Institute, 119992 Moscow, Russia\\
$^{2}$ Space Research Institute, Russian Academy of Sciences, Profsoyuznaya 84/32, 117997 Moscow, Russia\\
$^{3}$ Moscow Institute of Physics and Technology, Institutskyi per. 9, 141700 Dolgoprudnyi, Moscow Region, Russia\\
$^{4}$Tuorla observatory, Department of Physics and Astronomy, University of Turku, V\"ais\"al\"antie 20, 21500 Piikki\"o, Finland \\
}

\begin{document}

\date{Accepted .... Received ...}

\pagerange{\pageref{firstpage}--\pageref{lastpage}} \pubyear{2014}

\maketitle

\label{firstpage}

\begin{abstract}

\noindent We analyze spin-up/spin-down of the neutron star in Be X-ray binary system GX\,304-1 observed
by \textit{Swift}/XRT and \textit{Fermi}/GBM instruments in the period of the 
source activity from April 2010 to January 2013 
and discuss possible mechanisms of angular momentum transfer to/from the neutron star. 
We argue that the neutron star spin-down at quiescent states of the source 
with an X-ray luminosity of
$L_x\sim 10^{34}$~erg s$^{-1}$ between a series of Type I outbursts and 
spin-up during the outbursts can be explained by quasi-spherical
settling accretion onto the neutron star. The outbursts occur near the neutron star 
periastron passages where the density is enhanced due to the presence
of an equatorial Be-disc tilted to the orbital plane.  
We also propose an explanation to the counterintuitive smaller spin-up rate observed at
higher luminosity in a double-peak Type I outburst due to lower value of the specific
angular momentum of matter captured from the quasi-spherical wind from the Be-star by the
neutron star moving in an elliptical orbit \textbf{with eccentricity $e\gtrsim 0.5$}.

\end{abstract}


\section{Introduction}

In the early 1970s, the phenomenon of pulsating X-ray source due to accretion of matter
onto a magnetized rotating neutron star in a binary system was predicted
\citep[e.g.][]{Shvartsman71} and  discovered by UHURU satellite \citep{Giacconi_al71}.
Over the following decades a lot of observational properties of X-ray pulsars have been
established, including measurements of spin-up and spin-down rates from accurate timing
analysis \citep[see][for a review]{bil97} and measurements of cyclotron resonance features
in their X-ray spectra \citep[see][for a recent summary]{cab12}. The latter
suggests the presence of $10^{12}-10^{13}$~G magnetic fields near the surfaces of accreting neutron stars.
The measurements of spin-up/spin-down rate in X-ray pulsars reflect the torques applied to
the neutron star by accreting matter and can be served as a tool to investigate the interaction of
accreting plasma with magnetospheres of rotating neutron stars.

Most of the observed X-ray pulsars (XPSRs) recide in massive binary systems, with early-type optical
OB- or Be-companions, and belongs to the class of high-mass X-ray binaries (HMXB)
\citep[see, e.g.,][for a recent review]{lut09,lut13}. Formation of
HMXBs naturally follows from evolution of massive binary stars \citep{Bhattacharya_vandenHeuvel91}.
Accretion onto a neutron star (NS) in such a system occurs from a non-stationary stellar wind of
massive optical companions. If the specific angular momentum of gravitationally captured matter
is sufficiently high, an accretion disc can be formed around the neutron star, otherwise accretion
can proceed quasi-spherically (see e.g. \citealt{Burnard_al83} and \citealt{shak12,shak14a} for
recent discussion of different regimes of a quasi-spherical accretion onto NSs).


Measurements of pulse period variations in such systems can provide clues
for understanding different processes responsible for the angular
momentum transfer, the structure of the accretion flows in binary systems, their dependence on
system's parameters, X-ray luminosity, etc. This fact was recognized practically immediately
after the  discovery of X-ray pulsars \citep{schre72} and first observations of their
pulse period variations \citep[see, e.g.][]{gia73,fab77,beck78}. Different physical mechanisms for
torques acting on the magnetized neutron star in X-ray pulsars have been proposed to explain
these variations \citep[see, e.g.][]{gho79,lov95,wang95,rap04,kluzh07,shak13a}.  Observational data
suggest that different mechanisms can operate in different objects, and even in one and the
same object under different conditions.

Strong progress in observational X-ray astronomy in the last decade with modern space
observatories and instruments, especially with all-sky X-ray facilities, like \textit{Swift,
MAXI, Fermi}/GBM, \textit{RXTE}/ASM allowed us to monitor states of many X-ray sources and measure their pulse
periods with a high precision. These data can be used to test and verify different theories
of the angular momentum transfer in binary systems. Among X-ray pulsars, binary systems with
Be-companions form a subclass showing significant variations of the source luminosity and
pulse period. The stellar wind from rapidly rotating Be-stars is strongly asymmetric and
frequently forms equatorial excretion disc. In addition, such systems typically have a large
orbital eccentricity resulting from relatively recent supernova explosion. The presence of
these two features (equatorial disc and large eccentricity) enables the matter to accrete
onto the neutron star directly from the equatorial Be-star disc during its orbital motion
near the periastron passage. The enhanced accretion is accompanied by a short (compared to
the orbital period) increase in the observed X-ray flux by several orders of magnitude
(typically up to $\sim 10^{37}$~erg~s$^{-1}$), which is identified as a Type I outburst. In
addition to such events, there are Type II outbursts, which are due to a non-stationary
increase of amount of matter in the circumstellar disc around the Be-star, which can occur
at any orbital phase. The duration of Type II outbursts varies from weeks to months, during
which the source X-ray luminosity can reach the Eddington limit ($\sim10^{38}$ erg
s$^{-1}$). More details about X-ray pulsars in Be-systems can be found in the recent
review of \citet{reig11}.

During such outbursts, in addition to the X-ray luminosity increase, the angular momentum from
the equatorial disc around the Be-star is transferred to the compact object, leading to 
the neutron star spin-up. However, the angular momentum is supplied only when the neutron
star is sufficiently close to the optical companion, and after some time the accretion rate
starts decreasing and even can vanish altogether. From that moment on the gradual spin-down
of the neutron star is observed as a rule (see, e.g., site of the \textit{Fermi/GBM} data
http://gammaray.nsstc.nasa.gov/gbm/science/pulsars/). Obviously, the observed
characteristics of X-ray pulsars, including outbursts and pulse period variations, and
the corresponding mechanisms of the angular momentum transfer depend upon properties of the
binary system and the neutron star itself, including the orbital period, eccentricity,
magnetic field and spin frequency of the neutron star, etc.

In this paper, we examine the spin evolution of the X-ray pulsar
GX\,304-1 with the aim to understand the angular momentum transfer mechanism to
the neutron star.  GX\,304-1 (other names are 4U\,1258-61, 2S\,1258-613) is the
X-ray pulsar with a period of $\simeq272$ sec \citep{huck77,mcc77}, which is
located in the direction of Coalsack Nebula, in a highly optically opaque
region of the sky. GX\,304-1 has been identified with a Be-star system
\citep{mas78}, which is located at a distance of $d\simeq2.4$ kpc \citep{par80}.
Using observations by the \mbox{\textit{Vela 5B}} satellite, \citet{pred83}
 discovered a $132.5$-day periodicity of outbursts due to the orbital motion of
the neutron star. After approximately 30 years of quiescence, at the end of 2009
the source renewed the outburst activity that lasted until the beginning of
2013. Observations of the \textit{RXTE} and \textit{Suzaku} satellites during the
outburst in August 2010 allowed \citet{yam11} to  discover the cyclotron resonance
scattering feature in the source spectrum at around $\simeq54$ keV and to
estimate the neutron star surface magnetic field $B\simeq4.7\times10^{12}$~G.
Timing and spectral analysis of RXTE/PCA observations of 
this outburst by \cite{devasia11} discovered a QPO feature at frequency 
$\sim 0.125$~Hz presented in the power-density spectrum 
with an rms amplitude increasing from $\sim 3\%$ at 7~keV to $\sim 9\%$ at 40 keV, 
which was interpreted as evidence for an accretion disc existing during the outburst.
An analysis of 
the INTEGRAL observations of the source during its outburst in January - February 2012
\citep{kloch12} revealed a positive correlation of the cyclotron line centroid energy 
with X-ray flux variations, suggesting the local sub-Eddington regime of accretion near the NS surface \citep{staub07}. 
Thus, observations of the X-ray pulsar GX\,304-1 offer 
good opportunity to study different possible mechanisms of
the neutron star spin-up/spin-down.

\section{Data reduction and observational results}

In the period from April 2010 to January 2013 GX\,304-1 demonstrated a series Type I outbursts corresponding to periastron passages with the binary orbital period $P_b\approx 132.5$~days. During these
outbursts GX\,304-1 was monitored in hard X-rays (15-50 keV) by the \textit{Swift}/BAT instrument
(\citealt{krimm13}, \mbox{http://swift.gsfc.nasa.gov/results/transients/weak/GX304-1}); it was also ob\-se\-rved several dozen times
with the \textit{Swift}/XRT telescope in soft X-rays $<10$~keV \citep{XRT}. Variations of its pulse period
have been recorded with the \textit{Fermi}/GBM monitor
(http://gammaray.msfc.nasa.gov/gbm/science/pulsars/lightcurves/ gx304m1.html). In the present paper,
we made use of all of these data to examine spin-up/spin-down behavior of the neutron star in GX\,304-1.

\begin{figure}
\centering
\includegraphics[width=1.0\columnwidth,bb=15 160 582 708, clip]{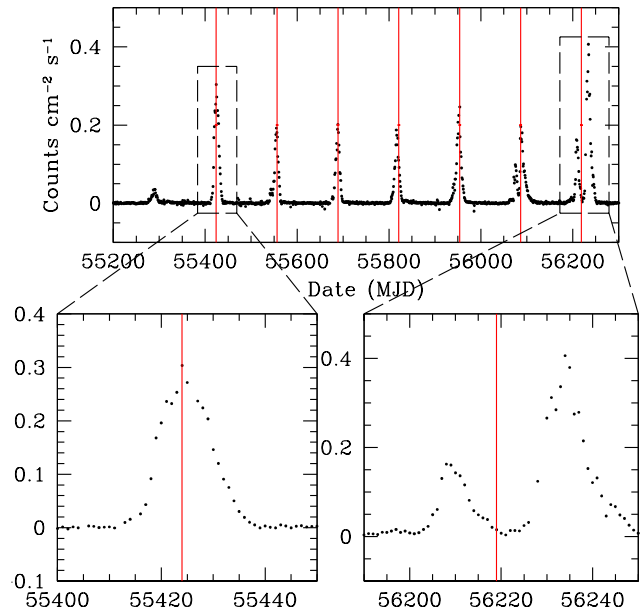}
\caption{The observed X-ray flux from GX\,304-1 measured by the \textit{Swift}/BAT telescope.
Vertical red lines indicate the moments of the periastron passages (for outbursts \#1-\#7). 
\label{fig1}}
\end{figure}
\begin{figure}
\centering
\includegraphics[width=1.0\columnwidth,bb=39 321 566 708, clip]{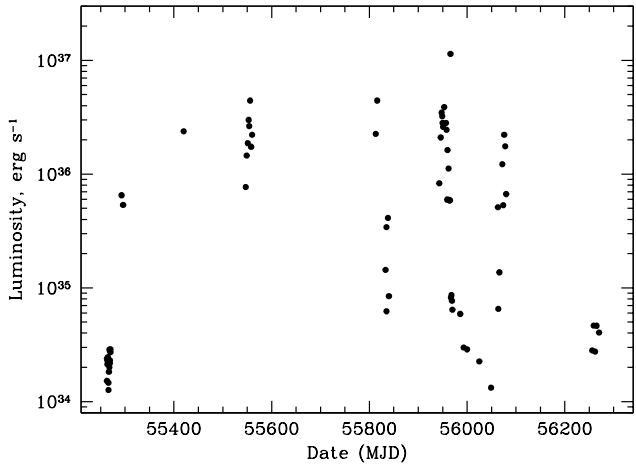}
\caption{The observed luminosity of GX\,304-1 measured by the \textit{Swift}/XRT instrument
in the 2-10 keV energy band. Note a big power difference between the  outbursts and
quiescent states.
\label{fig2}}
\end{figure}

The source hard X-ray light curve measured by the \textit{Swift}/BAT telescope 
is shown in Fig.\ref{fig1}. Seven giant outbursts (below referred to as \#1-\#7), 
where the source intensity reached about 2 Crabs
in the $15-50$ keV energy band, are clearly seen in the light curve. Note here that
the pulsar's X-ray luminosity in the outbursts increases by almost three orders 
of magnitude in $\sim 10$
days, suggesting a very quick and strong increase in accretion rate onto the neutron star.
Outbursts profiles from \#1 to \#5 are very similar --
they have a nearly symmetric shape and their maxima are separated by the orbital period (indicated
by vertical red lines in Fig.\ref{fig1}). One of such outbursts (\#1) is presented in the left insertion in the
figure. At the same time, the profiles of two outbursts (\#6 and \#7) are completely different --
they have a double-peak structure with a lower precursor and dominating second peak; in addition, the orbital phase
corresponding to maxima of outbursts \#1-\#5, here occurs at the minima between the outburst peaks (see the right
insertion in Fig.\ref{fig1}). Possible reasons of such drastic changes in the outburst profile
may be connected with the evolution and changes in the equatorial disc around the Be-star and will
be  discussed below.

The corresponding behavior of the source luminosity in the $2-10$ keV energy band
according to the XRT telescope data is presented in Fig. \ref{fig2} and Table 1. 
The reduction of data obtained
by the XRT telescope was done using the standard software {\sc FTOOLS v6.15}.
When no XRT observations were available, 
to estimate a peak X-ray flux in the 2-10 keV we used the \textit{Swift}/BAT 15-50~keV flux  
and assumed the spectral shape $E^{-\Gamma}\exp(-E/E_f)$ 
with parameters $\Gamma=1$, $E_f=17$~keV from \cite{kloch12}.

As mentioned above, the source intensity increased by about three orders of magnitude during
outbursts in comparison with the quiescent state. It is necessary to note that in contrast to
the neutron star spin-up, processes of the neutron star spin-down in X-ray binaries are still
poorly understood. In wind-fed X-ray pulsars, the spin-down torque depends on the source luminosity
and other parameters \citep{shak12, shak13a, postnov14}.
It is seen from Fig. \ref{fig2} and Table 1
that in the low states between outbursts the $2-10$~keV luminosity
of GX\,304-1 drops down to a $few \times10^{34}$ \ergs. Note that GX\,304-1
was observed with the \textit{Swift}/XRT telescope starting from April 2005, i.e. long before the beginning of its
outburst activity, and during these observations the source luminosity stayed virtually at the same low
level. Thus we can conclude that the lowest observed (at the moment) luminosity for the source
\mbox{GX\,304-1} is about $(1-2)\times 10^{34}$ \ergs\ in the 2-10 energy band.\footnote{Note that
the total (bolometric) luminosity of X-ray pulsars can be about three times as high \citep[see][for details]{fil05},
but the correction factor depends on the broadband spectrum of the source. Taking into account
uncertainties in the distance measurements to GX\,304-1 \citep{par80,men81}, in our
estimations we use the source luminosity as derived from the XRT fluxes.}

\begin{figure}

\centering
\includegraphics[width=1.0\columnwidth,bb=20 160 582 708, clip]{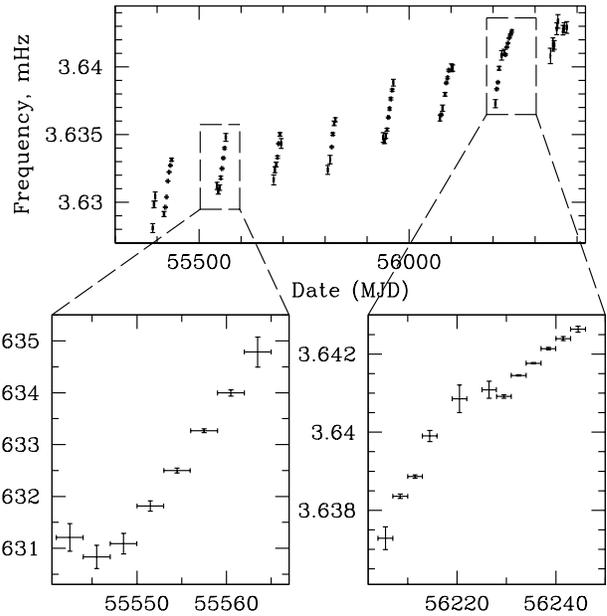}
\caption{Spin frequency evolution of GX\,304-1 as measured by the \textit{Fermi}/GBM monitor.
Note that the double-peak outburst \#7 is characterized by two different spin-up
rates. \label{fig3}}
\end{figure}

The evolution of the neutron star spin frequency
in GX\,304-1 in the period from August 2010 to March 2013 as  
obtained from the \textit{Fermi}/GBM monitor data is presented in Fig.\ref{fig3}.
As seen from  Figs.\ref{fig1} and \ref{fig3}, 
the neutron star exhibits regular spin-up episodes during Type I outbursts 
when the source flux is high enough 
for X-ray pulsations to be detected. The neutron star spin-up behaviour is different
during different outbursts -- typically it is nearly constant for single-peak Type I outbursts
(the left insertion in Fig.\ref{fig3} for outburst \#1) and have an obvious break for the double-peak outburst \#7 (the right insertion in Fig.\ref{fig3}).

To determine the spin-up rate in the outbursts, 
the linear trends of the spin frequency changes were approximated with
the least squares method. Thanks to high quality of the \textit{Fermi}/GBM data these measurements
can be done with good accuracy (see Table\,\ref{tab1}). 
It is seen from Table 1 that the spin-up rate
during single-peak Type I outbursts was 
approximately the same $\dot\omega^*_{su}\simeq 25\times 10^{-8}$~Hz~d$^{-1}$.
In contrast, the double-peak outburst \#7 demonstrates two different values of the spin-up rate:
while during the first, weaker, peak the spin-up rate was approximately the same (within the errors)
as during previous single-peak outbursts, it was about three times as low during the second, higher,
peak (see Fig.\,\ref{fig3} and Table\,\ref{tab1}).

Spin-down rate estimates during the low states 
of the source between the outbursts are not so obvious
and straightforward, because no X-ray pulsations were detected by the \textit{Fermi}/GBM monitor.
We can only compare the neutron star spin frequency at the end of the spin-up phase of
one outburst with that measured at the beginning of the next outburst.

It is interesting to note that at the beginning of most of the outbursts short episodes
of the continuing spin-down can be seen. This can be easily understood, because in order
to change the neutron star rotation from spin-down to spin-up 
a certain amount of angular momentum should be transferred to
overcome the neutron star inertia.

Results of the spin-up/spin-down measurements in GX\,304-1 are summarized in Table\,\ref{tab1}. Two main
facts can be learned from this table: the spin-down rates between outbursts are the same (within errors)
and are about one order of magnitude (by the absolute value) lower  than the neutron star spin-up rates during the
outbursts. These facts are 
 discussed below in the frame of quasi-spherical settling accretion onto the
neutron star in GX\,304-1 from the stellar wind of the Be-star.

\begin{table}
\noindent
\centering
\caption{Spin-down rate between Type I outbursts and spin-up rates during outbursts in GX\,304-1.}\label{tab1}
\centering
\vspace{1mm}
\small{
\begin{tabular}{c|c|c|c}
\hline\hline
Outburst \# & Period &  $\dot{\omega}$ & Peak flux (2-10 keV)\\
& MJD &  ($10^{-8}$ Hz~d$^{-1}$) & erg s$^{-1}$~cm$^{-2}$ \\
\hline
\multicolumn{4}{c}{Spin-down}\\
\hline
1 & 55434-55546 & $-2.08\pm0.29$ & \\[1mm]
\vspace{1mm}
2 & 55563-55677 & $-2.74\pm0.58$ & \\
\vspace{1mm}
3 & 55692-55806 & $-2.3\pm0.41$ & \\
\vspace{1mm}
4 & 55824-55941 & $-1.4\pm0.27$ & $9.0\times10^{-11}$\\
\vspace{1mm}
5 & 55962-56073 & $-2.34\pm0.42$ & $1.9\times10^{-11}$\\
\vspace{1mm}
6 & 56101-56205 & $-2.53\pm0.55$ &\\
\vspace{1mm}
7 & 56244-56336 & $-1.98\pm0.7$ & $4.0\times10^{-11}$\\
\hline
\multicolumn{4}{c}{Spin-up}\\
\hline
1 & 55416-55434 & $23.8\pm3.4$ & $1.7 \times10^{-8}$ \\[1mm]
\vspace{1mm}
2 & 55548-55563 & $24.6\pm0.47$ & $1.1 \times10^{-8}$ \\
\vspace{1mm}
3 & 55677-55692 & $22\pm2.8$ & $1.1\times10^{-8}$ \\
\vspace{1mm}
4 & 55811-55821 & $28.1\pm2.9$ & $1.1\times10^{-8}$ \\
\vspace{1mm}
5 & 55944-55962 & $22.6\pm1.4$ & $1.4\times10^{-8}$ \\
\vspace{1mm}
6 & 56073-56094 & $17.6\pm1.8$ & $1.1\times10^{-8}$ \\
\vspace{1mm}
7$^{a}$ & 56205-56214 & $27.9\pm3.7$ & $0.9\times10^{-8}$ \\
\vspace{1mm}
7$^{b}$ & 56229-56244 & $11.3\pm1.6$ & $2.3\times10^{-8}$ \\
\hline
\end{tabular}}
\medskip
\begin{tabular}{ll}
    $^{a}$  & For the precursor of outburst \#7 (see Fig.\ref{fig3})\\
    $^{b}$  & For the second giant peak of outburst \#7 (see Fig.\ref{fig3})\\
    \end{tabular}
    \medskip
\end{table}

\section{Quasi-spherical accretion in GX\,304-1}
\label{s:qs}

The observed spin-up/spin-down behavior of GX\,304-1 can be readily explained in the frame of
theory of quasi-spherical accretion from stellar wind of the optical companion. The settling
accretion is expected to occur in wind-fed accreting neutron stars with sufficiently slow spin periods 
at X-ray luminosities $L_x<4\times 10^{36}$~\ergs. The basic difference of this regime 
from the classical Bondi-Hoyle case is that the matter flow to the neutron star 
is controlled by the development of the Rayleigh-Taylor 
instability in a boundary layer above the magnetosphere, which depends on the 
cooling time $t_{cool}$ of accreting plasma. The mean plasma radial infall velocity in this regime 
is less than the free-fall velocity: $u_r=u_{ff}f(u)$, where $f(u)\sim (t_{cool}/t_{ff})^{1/3}<0.5$
($t_{ff}=R/u_{ff}$ is the free-fall time). Correspodingly, 
the accretion rate onto the neutron star that determines the observed X-ray luminosity is 
$\dot M\simeq \dot M_B f(u)$, where $\dot M_B$ is the classical Bondi-Hoyle rate. 
See \cite{shak12, shak13b, shak14a} for
the detailed derivation,  discussion and applications to spin-up/spin-down of slowly rotating
low-luminosity X-ray pulsars and \cite{shak14b} for explanation of bright flares in Supergiant
Fast X-ray Transients.

The X-ray luminosity of GX\,304-1 stays most of the time at a low level of a
few~$\times 10^{34}$~\ergs, and the peak luminosities during the 
most of Type I outbursts is less
than $10^{37}$~\ergs (see Fig.\,\ref{fig1},\ref{fig2}), which favors the settling accretion regime.
Next, the observed spin period $P_*\approx 275$~s of the neutron star in GX\,304-1 with the
standard surface magnetic field value, as inferred from cyclotron line measurements,
$B \simeq 4.7\times 10^{12}$~G, is consistent with the expected equilibrium period of
quasi-spherically accreting neutron stars in the regime of settling accretion

\beq{Peq}
P_{eq}^*\approx 940[\mathrm{s}]\mu_{30}^{12/11}\myfrac{P_b}{10\mathrm{d}}
\dot M_{16}^{-4/11} v_8^{4}\,,
\eeq

\noindent where $\mu_{30}\equiv \mu/10^{30}[\mathrm{G\,cm}^{3}]$ is the neutron star dipole magnetic
moment related to the surface equatorial dipole magnetic field as $\mu=BR^3/2$ ($R$ is the
neutron star radius assumed to be 10 km),  $\dot M_{16}\equiv \dot M/10^{16}[\mathrm{g\,s}^{-1}]$
is the accretion rate onto the neutron star related to the accretion X-ray luminosity as 
$L=0.1\dot Mc^2$, $P_b$ is the binary orbital period and $v_8\equiv v/10^8[\mathrm{cm \,s}^{-1}]$ is the
characteristic stellar wind velocity. As stressed in \citet{shak14a}, due to very strong
dependence upon the stellar wind velocity ($P_{eq}^*\propto v^{4}$), this formula, when the neutron star magnetic
field $\mu$ is known, can be rather used to estimate the velocity of the stellar wind captured by the
neutron star:

\beq{v8}
v_8\approx 0.57 \dot M_{16}^{1/11}\mu_{30}^{-3/11}\myfrac{P_{eq}^*/100\mathrm{s}}{P_b/10\mathrm{d}}^{1/4}
\eeq

Substituting for GX\,304-1 $\mu_{30}\approx 2.35$\footnote{Using 
$B=10^{12}[\hbox{G}]E_{cyc}/(11.6 \hbox{keV})$ and neglecting 
gravitational redshift.}, $P_*=275$~s,  $P_b=132.5$~d and $\dot M_{16}\sim
0.02$ (for the low-state, where the source spends most of the time; this, however, is not very
important in view of very weak dependence on $\dot M$ in \Eq{v8}), we find $v_8\sim 0.2$. This low
wind velocity is typical for quasi-spherical winds observed in Be-stars \citep{wat88}.
We also note that the QPOs at $\sim 0.125$~Hz reported by \cite{devasia11} during 
outburst \#1 may be due to the accretion rate variations with typical free-fall time
from the Alfv\'en radius $\sim 3\times 10^9$~cm in the quasi-spherical accretion case.
Indeed, usually the presence of QPOs is interpreted as an indication of the accretion disc. 
For example, in the popular beat-frequency model \citep{alp85}, the QPO frequency is treated as the 
beat frequency between the magnetospheric rotation and the matter rotation at the inner disc
radius. In the quasi-spherical case, there 
is the characteristic time related to the Rayleigh-Taylor instability  
at the magnetospheric boundary, which is of the order of the free-fall time of matter 
at the magnetosphere. Therefore, one can expect 
the characteristic variability in X-ray flux with this time, i.e. 
at the frequency corresponding to the free-fall time and its harmonics. 
QPOs arising due to matter entering the rotating neutron star magnetosphere
via instabilities were discussed in \cite{jern00}.

The characteristic parameters and radii pertinent to neutron star spin-up/spin-down in
GX\,304-1 are summarized in Table 2.

\begin{table*}
\label{t:radii}
\caption{Parameters of GX\,304-1}
$$
\begin{array}{rl}
\hline
\hline
\multicolumn{2}{c}{\mathrm{Observed}}\\
\hline
 \hbox{Pulsar period}, P_* & 275\,\mathrm{s}\\
 \hbox{Binary period}, P_b & 132.5\,\mathrm{d}\\
\hbox{NS magnetic moment},\mu & 2.35\times 10^{30}\mathrm{G\,cm}^3\\
\hbox{Low-state accr. rate}, \dot M_{low} & 2\times 10^{14}\mathrm{g\,s}^{-1}\\
\hbox{Accr. rate in outbursts}, \dot M_{outb} & 3-7\times 10^{16}\mathrm{g\,s}^{-1}\\
\hbox{Mean spin-up in outbursts},
\dot\omega^*_{su} & -2.5\times 10^{-8}\,\mathrm{Hz\, d}^{-1}\\
\hbox{Mean spin-down btw. outbursts},
\dot\omega^*_{sd} &  25 \times 10^{-8}\,\mathrm{Hz\, d}^{-1}\\
\hline
\hline
\multicolumn{2}{c}{\mathrm{Derived}}\\
\hline
\hbox{Corotation radius}\, R_c & \myfrac{GM P_*^2}{4\pi^2}^{1/3}\approx 7.3\times 10^9\,\hbox{cm}  \\
\hbox{Alfv\'en radius, q.-s.}\, R_{A,qs} & 1.35\times 10^9[\hbox{cm}] \myfrac{\mu_{30}^3}{\dot M_{16}}^{2/11}\\
\hbox{Alfv\'en radius,  disc}\, R_{A, disc} & \approx 5.9\times 10^8[\hbox{cm}]\myfrac{\alpha_{SS}\mu_{30}^2}{\dot M_{16}}^{2/7}
 \\
\hbox{Wind velocity}, v & \approx 200\,\mathrm{km\,s}^{-1}\\
\Pi_{su/sd} & 4.6 \\
\hline
\end{array}
$$
\end{table*}

\subsection{Spin-down between Type I outbursts}

In the frame of the quasi-spherical settling accretion theory \cite{shak12, shak14a},
the equilibrium X-ray luminosity at which accretion torques acting on the neutron star with parameters as in GX\,304-1 vanishes, is $L_{eq}\sim 10^{36}$~\ergs. At lower X-ray luminosities,
when $\dot M_{low}\ll \dot M_{eq}\approx 10^{16}$~g s$^{-1}$, the spin-down rate is

\beq{sd}
\dot\omega^*_{sd}\approx -10^{-8}\,[\mathrm{Hz\,d}^{-1}] \Pi_{sd} \mu_{30}^{13/11}\dot M_{16,low}^{3/11}\myfrac{P_*}{100\,\mathrm{s}}^{-1}
\eeq

\noindent where $\Pi_{sd}=(1-z/Z)\tilde K K_1 K_3\zeta^{-3/11}$ is a combination of dimensionless parameters of the theory (in notations of Eq. (42) in \cite{shak14a}). The physical sense of these parameters is as follows: $\tilde K\sim 1$ is a geometrical factor
due to  integrating the magnetic  
torques over the magnetospheric surface, $K_1\sim 1$ is a factor 
that takes into account the difference of the 
realistic magnetospheric shape from the sphere and the characteristic scale of 
turbulent motions near
the magnetosphere (in units of the Alfv\'en radius $R_A$), $K_3$ is the factor that determines the
coupling of the rotating matter with the magnetosphere and takes into account 
the structure of the logarithmic boundary layer above the magnetosphere, $\zeta\lesssim 1$ is the dimensionless thickness of the boundary layer (in units of $R_A$), $0<z<2/3$ is 
the dimensionless specific torque applied to the neutron star in units of $\dot M R_A^2\omega^*$ 
due to matter falling from different parts of the magnetosphere, $Z\sim \tilde K K_1 K_3/f(u)$ is the dimensionless coupling coefficient describing of the angular momentum transfer to the magnetosphere in the settling accretion regime.
This combination of parameters should not be strongly different in different objects (see \citealt{shak14a, postnov14}
for more details). For parameters of GX\,304-1 and $\dot M_{16,low}\simeq 0.02$,  \Eq{sd} yields the observed value
$\dot\omega_{sd}^*\simeq -2.5\times 10^{-8}$~Hz d$^{-1}$ for $\Pi_{sd}\approx 4.6$. Note that this value
is reasonably close to the value of $\Pi\simeq 9$ derived from the independent analysis of equilibrium wind-accreting X-ray  
pulsars Vela\,X-1 and GX\,301-2 \citep{shak12, shak14a}.

\subsection{Spin-up during Type I outbursts}

As we  discussed in Section 2, the spin-up episodes are observed during Type I  outbursts of
GX\,304-1, which regularly occurred near the orbital periastron passages (see Fig. 3 and Table
1). In all but the last outburst we analyze here (\#7), the spin-up rate was found to be approximately the same,
$\dot\omega^*_{su}\simeq -25 \times 10^{-8}\,\mathrm{Hz\, d}^{-1}$.
In the theory of settling quasi-spherical accretion (eqs.
(40-41) in \citealt{shak14a}), the spin-up rate  (neglecting spin-down torque, which is
justified at $\dot M_{16}\sim 1$ for the GX\,304-1 parameters) reads:

\beq{su}
\dot\omega^*_{su}\simeq 10^{-9}\,[\mathrm{Hz\, d}^{-1}]\Pi_{su}\mu_{30}^{1/11}v_8^{-4}\myfrac{P_b}{10\,\mathrm{d}}^{-1}\dot M_{16}^{7/11}
\eeq

\noindent where the dimensionless parameter of the theory $\Pi_{su}=\tilde K K_1 K_3\zeta^{-7/11} \approx \Pi_{sd}$ (see Eq. (41) in \cite{shak14a}).
Taking $\Pi_{su}\simeq 4.6$ as inferred from the analysis of the spin-down between the outbursts
(see the previous paragraph), we obtain $\dot\omega_{su}^*\approx 25\times 10^{-8}\,\mathrm{Hz\, d}^{-1}$, as actually observed. Therefore, we conclude that the observed spin-up rate in Type I outbursts \# 1-\#6
and the spin-down rate at low states between the outbursts are in agreement with the settling accretion
theory predictions for GX\,304-1.

\subsection{Double-peak outburst \#7}

The double-peak outbursts \#6 and \#7 (see Fig.\ref{fig3}, right inset for \#7) 
make the special case. Indeed, unlike
previous Type I outbursts, these outbursts show a double-peak structure, mostly pronounced in outburst \#7.
This outburst consists of a fainter precursor that peaked $\sim 10$ days before the periastron passage and lasted
$\sim 10$ days, and was followed by a brighter main flare that peaked $\sim 15$~days
after the periastron passage. 
The most enigmatic feature is that the spin-up rate during the
weaker precursor was $\sim 3$ times as high as that of the brighter second flare.
This is counterintuitive, since the spin-up torque in any model is proportional to $\dot M$
(e.g. $\sim \dot M^{6/7}$ for the standard  disc accretion \citep{gho79} and $\propto \dot M^{7/11}$ for the
settling quasi-spherical accretion \citep{shak12}).

\begin{figure}
\centering
\includegraphics[width=1.0\columnwidth,bb=60 0 1499 1126, clip]{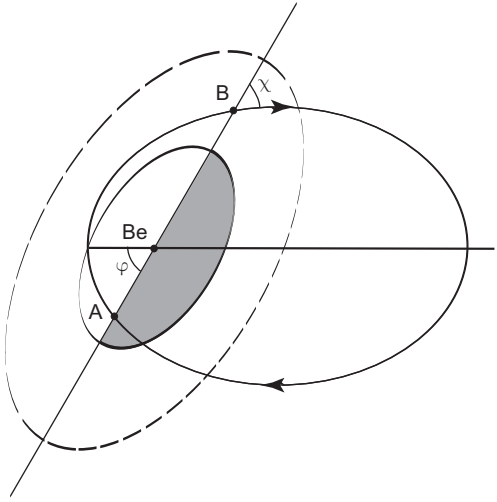}
\caption{Schematics of the neutron star orbit around Be-star surrounded by a tilted equatorial  disc
(face-on view). The neutron star first meets the  disc node line at point A and leaves the  disc at
point B. In so far as the Be-disc radius is smaller than or comparable to the periastron distance, a
single-peak Type I outburst is expected to occur near the periastron passage. As the radius of the
Be-disc increases (the dashed ellipse), the neutron star meets the  disc at point A and leaves it at
point B.  
\label{fig4}}
\end{figure}

A plausible explanation can be as follows. The stellar wind from rapidly rotating Be-star is highly asymmetric and  can form an excretion disc in the equatorial plane of the Be-star
(see \cite{thomas79} for spectroscopic evidence of an equatorial disc around Be-star 
in GX\,304-1). The Be-disc can be inclined to the
orbital plane of the binary system due to, for example, non-zero kick velocity acquired by the
neutron star during the supernova explosion. In this case the nodal line of the  disc and the
orbital plane may not be perpendicular to the major semi-axis line, i.e. the orbit of the
neutron star around the Be-star can enter the  disc at smaller radius ($r_A$) and go out of the
disc at larger radius $r_B>r_A$, as depicted in Fig. \ref{fig4}. 
The precursor initiated by the stellar wind density enhancement when neutron star 
approaches point A can still be at the stage of setlling accretion (the peak luminosity 
is about $6\times 10^{36}$~\ergs, which is marginal for this regime with $f(u)\sim 0.5$
\citep{shak12}, allowing for distance 
uncertainties to the source), so the neutron star 
spin-up properties should be similar to the previous 
outbursts. The main flare after the periastron at point B is two times as bright, i.e. must be 
in the free-fall Bondi regime with $f(u)=1$. In this case, the neutron star spin-up rate is 
entirely determined by the specific angular momentum of the captured stellar wind matter, 
which should change along the non-circular orbit.

\begin{figure}
\hskip -4cm
\includegraphics[width=\columnwidth]{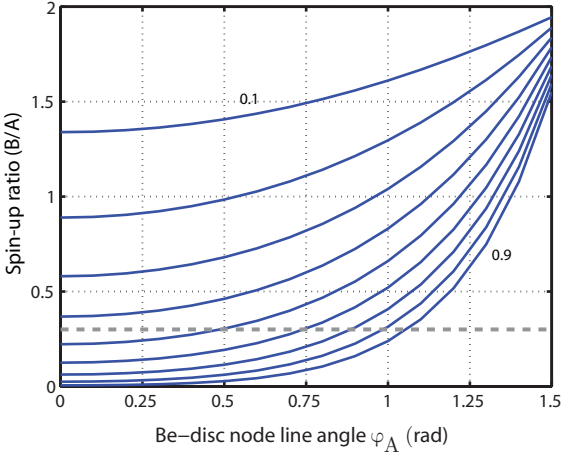}
\caption{Ratio of the pulsar spin-up rates at the neutron star
crossing the Be-disc node line (points B and A in Fig. 4), \Eq{A/B}, as a function of the 
angle $\varphi_A$ for different orbital eccentricities (0.1 to 0.9, the solid lines from top to bottom). The dashed gray line corresponds to the observed ratio 0.3 in the main (B) and precursor (B)flares in outburst \#7.  
\label{f:fig5}}
\end{figure}

Consider a neutron star with mass $M_{NS}$ 
in elliptical orbit around the central star with mass $M\gg M_{NS}$, so that we can 
treat the star $M$ to be at rest.
The orbital eccentricity $e$ of GX\,304-1 is unknown, but likely can be substantial. 
The position of the neutron star in the orbit is characterized by the true anomaly $\varphi$
(see Fig. 4), so that the radius-vector is 
\beq{e:r}
r=\frac{p}{1+e\cos\varphi}
\eeq
and $p=a(1-e^2)$ is the semilatus rectum. The specific angular momentum of the neutron star 
is constant along the orbit and is $|\bm{j}|=\sqrt{GMp}=|{\bm v}||\bm{r}|\sin\chi$, where 
$\chi$ is the angle beween the neutron star orbital velocity and the radius-vector (see Fig. 4).
Assuming the neutron star moving in the spherically-symmetric wind from the central Be-star, the
specific angular momentum captured from this wind is $\eta \Omega R_B^2$, where 
$R_B=2GM_{NS}/(v_w^2+v^2)$ is the Bondi radius, $\eta\sim 1/4$ is the numerical coefficient \citep{is75} and  
the instantaneous angular
velocity is determined by the wind tangential velocity relative to the neutron star, i.e. 
\beq{e:omega}
\Omega=\frac{v_t}{r}=\frac{v\sin\chi}{r}=\frac{j}{r^2}\,.
\eeq
Assuming $v_w\gg v$ and substituting for $j$ and $r$ from \Eq{e:r} into \Eq{e:omega}, we 
find the spin-up torque 
\beq{e:su}
I\dot\omega^*_{su}=\dot M j=\frac{1}{4}\dot MR_B^2\sqrt{\frac{GM}{p^3}}(1+e\cos\varphi)^2\,.
\eeq
(Here $I\approx 10^{45}$~g~cm$^2$ is the neutron star moment of inertia). 
Therefore, 
\beq{}
\frac{\dot\omega^*_{su}(B)}{\dot\omega^*_{su}(A)}=\frac{\dot M(B)}{\dot M(A)}\myfrac{1+e\cos\varphi_B}{1+e\cos\varphi_A}^2\,.
\eeq
In our case the true anomaly of two flares (before and after periastron) 
is determined by the Be-disc node line, i.e. $\varphi_B=\varphi_A+\piup$, and neglecting wind velocity variations with radius
(i.e. assuming $v_w=const$), we obtain $\dot M(B)=\dot M_B$ (Bondi regime), $\dot M(A)=f(u)\dot M_B\sim 0.5 \dot M_B$ (settling regime), so that
\beq{A/B}
\frac{\dot\omega^*_{su}(B)}{\dot\omega^*_{su}(A)}\approx 2\myfrac{1-e\cos\varphi_A}{1+e\cos\varphi_A}^2\,.
\eeq
Clearly, it is not difficult to find the combination of the orbital eccentricity $e$ and 
the true anomaly $\varphi_A$ matching the observed spin-up ratio 
$\dot\omega^*_{su}(B)/\dot\omega^*_{su}(A)\sim 0.3$ \textbf{(see Fig. \ref{f:fig5}). The
acceptable angle $\varphi_A$ varies from $\sim 0.5$~rad to $\sim 1.1$~rad 
for orbital eccentricity ranging from $\sim 0.5$ to $0.9$, respectively.}
We stress that in this picture the 
role of the equatorial Be-disc is simply in enhancing the wind density when the neutron star
approaches the disc node line in its orbital motion. 
The specific angular momentum of the 
Be-disc matter at points A and B should be of the order of the Keplerian value $\sqrt{GMr}$, which is close to the orbital neutron star value. But under our assumption $v_w\gg v$ its addition to 
the captured spherical wind matter is not expected to be substantial.

Thus we can conclude that the model of quasi-spherical settling accretion is consistent with
observations of both spin-up episodes during the outbursts at periastron passages of the neutron
star and spin-down between the outbursts in slowly rotating wind-fed X-ray pulsar GX\,304-1. 
The first, weaker, flare (precursor) in the unusual double-peak outburst started before the periastron passage at
MJD 56205 can be due to continuing quasi-spherical accretion from the quasi-spherical 
low-velocity wind of Be-star, while the second, more powerful flare,
developed after the periastron passage which started at MJD 56229, 
is in the free-fall Bondi regime of quasi-spherical acretion. The lower
spin-up rate during the second flare is explained by smaller specific
angular momentum of matter captured from the quasi-spherical wind from the Be-star by the
neutron star moving in elliptical orbit.

Apparently, the
complicated behavior of the excretion  disc around Be-star may cause the observed switching
between accretion regimes near the periastron passage. We also note that the time separation between
the precursor and the giant flare in double-peak outbursts \#6 and \#7 increases (see Fig. 1), which may be due to
growing radius of the excretion  disc around the Be-star. The growing radius of the Be-disc
in principle could be checked by optical observations \citep{corbet86}. Unfortunately, no 
optical monitoring was reported during the source activity in 2009-2013.

\section{ discussion and conclusion}

Let us  discuss an alternative explanation to the observed spin-up/spin-down behavior in
GX\,304-1. First, assume that an accretion  disc is always present around the neutron star
(see e.g. SPH simulations \citep{hayazaki04}).
Observations of QPOs in outburst \#1 \citep{devasia11} may be an indication of the disc.
 During
giant outbursts the total accretion luminosity of the source can be as high as $10^{37}$~\ergs 
(see Table 1),
which can be uncomfortably high for the settling quasi-spherical accretion. The spin-up rate
during disc accretion can be written as
\beq{sud}
\dot\omega^*_{su,disc}\sim \dot M\sqrt{GMR_{disc}}/I\simeq 4.6\times 10^{-8}\,[\mathrm{Hz\,d}^{-1}]\dot M_{16}^{6/7}
\eeq
(here we neglected spin-down torques due to interaction of the magnetosphere with accretion
disc (see e.g. discussion of models in \citealt{wang96, kluzh07}), which is
justified at high accretion rates, 
and assumed $\alpha=0.1$ for the thin accretion disc in GX\,304-1).
Clearly, the plausible mass accretion rate onto the neutron star $\dot M_{16}\sim 7-8$ during Type I
outbursts in GX\,304-1 is sufficient to match the observed spin-up rate $\sim 25 \times
10^{-8}\,\mathrm{Hz\,d}^{-1}$ (see Table 1). A caveat in the disc spin-up at all times, however, is
that, as mentioned above, it is difficult to explain the \textit{lower} spin-up rate observed in
the second, \textit{more luminous}, flare of the double-peak outburst \#7. 

The persistent X-ray luminosity between outbursts clearly indicates that accretion does not stop during
the orbital motion of the neutron star even at the orbit apastron. While it is possible to produce a
spin-down torque during low-luminosity disc accretion, for example in models
\citep{wang96, kluzh07}, it is difficult to reconcile the observed
275-s neutron star period with these models, even by assuming that most of the time the
accretion rate is as low as $\dot M_{16}\sim 0.01$. Indeed, the equilibrium period in the
sophisticated model by \citet{kluzh07} is $P_{eq}\approx 1.14 [\mathrm{ms}] \dot
M_{18}^{-3/7}B_9^{6/7}\approx 83$~s for parameters of GX\,304-1 $B_9\approx 4.7\times 10^3$ and
$\dot M_{18}=10^{-4}$, more than three times as short as observed. (Incidentally, we stress here
the general problem of explaining the observed long periods of X-ray pulsars in BeXRB systems in
the frame of the standard disc accretion onto neutron stars with the standard magnetic field
about $10^{12}$~G; quasi-spherical settling wind accretion does not have this problem, see
\citealt{postnov14}).

Thus we conclude that the quasi-spherical settling accretion onto the slowly rotating neutron star
with standard magnetic field from stellar wind of the companion Be-star seems to be a likely
explanation to the observed spin-down between Type I outbursts in GX\,304-1. The angular momentum from
the neutron star is removed due to interaction of the magnetosphere with convective hot shell formed
from the matter with low angular momentum captured from the quasi-spherical low-velocity wind of the
Be-star. 
The Be-disc 
can be inclined to the binary orbital
plane due to the natal kick acquired by the neutron star during supernova explosion. When the radius
of the circumstellar disc is smaller or comparable with the periastron distance, a single-peak Type I
outburst occurs near the orbital periastron passage. When the radius of the Be-disc increases above
the orbital periastron distance, a double-peak outburst near the periastron can be produced, with the
second (post-periastron) peak being 
at the more effective quasi-spherical Bondi accretion stage from the 
stellar wind of Be-star. The lower
spin-up rate during the second flare can be explained by lower value of the specific
angular momentum of matter captured from the quasi-spherical wind from the Be-star by the
neutron star moving in elliptical orbit. Once the orbital parameters of GX\,304-1 will be 
determined, the orbital phase of the precursor and post-periastron flares in double-peak outbursts 
can be used to find the Be-disc node line angle. Further optical and X-ray observations of the system are encouraged to disentangle the complicated structure of the Be-star wind and to 
check the proposed model for X-ray outbursts.

\section*{Acknowledgements}

This work made use of data provided by \textit{Fermi}/GBM pulsar project
(http://gammaray.nsstc.nasa.gov/gbm/science/pulsars.html). We thank the 
anonymous referee for useful comments and important notes.
KP, NSh and AK acknowledge partial support by the RFBR grants 12-02-00186a, 14-02-00657a
and 14-02-91345,  and thank Max-Planck Institut
fuer Astrophysik (Garching) for hospitality. AM and AL acknowledge support by the grant of the
Russian Science Foundation 14-12-01287.

\label{lastpage}

\end{document}